%% file: emnlp2023.tex
\pdfoutput=1

\documentclass[11pt]{article}

\usepackage{EMNLP2023}
\usepackage{times}
\usepackage{latexsym}

\usepackage[T1]{fontenc}

\usepackage[utf8]{inputenc}
\usepackage{float}
\usepackage{microtype}

\usepackage{inconsolata}

\usepackage{fvextra} 

\usepackage{bbding}
\usepackage{colortbl}
\usepackage{graphicx}
\usepackage{changepage}
\usepackage{amsmath}
\usepackage{framed}
\usepackage{listings}
\usepackage{booktabs} 
\usepackage{pifont} 
\newcommand{\cmark}{\ding{51}} 
\newcommand{\xmark}{\ding{55}} 

\usepackage{chemformula}
\definecolor{codegray}{gray}{0.9}
\definecolor{codeblue}{rgb}{0.3,0.3,0.7}

\title{PyBench: Evaluating LLM Agent on various real-world coding tasks}

\author{
\textbf{Yaolun Zhang$^{1}$\thanks{~ Equal contribution.}$\;$\thanks{~ Work done during internship at ModelBest Inc.}$\;$, Yinxu Pan$^{2}$\footnotemark[1]$\;$, Yudong Wang$^{2}$\footnotemark[1]$\;$, Jie Cai$^{2}$$\;$} \\ 
$^{1}$School of Statistics, Renmin University of China \\
$^{2}$ModelBest Inc. \\
{\small \texttt{\{zhangyaolun5\}@ruc.edu.cn}} \\
{\small \texttt{\{panyinxu, wangyudong, caijie\}@modelbest.cn}} \\
}

\begin{document}
\maketitle
\input{abstract}

\input{introduction}

\input{related_works}

\input{bench}

\input{method}

\input{experiment}

\input{conclusion}

\input{limitations}

\bibliographystyle{acl_natbib}
\bibliography{reference}

\input{appendix}

\end{document}

%% file: abstract.tex
\begin{abstract}
The LLM Agent, equipped with a code interpreter, can automatically solve real-world coding tasks such as data analysis and image editing.
However, existing benchmarks primarily focus on either simplistic tasks, such as completing a few lines of code, or on extremely complex and specific tasks at the repository level, neither of which are representative of various daily coding tasks. 
To address this gap, we introduce \textbf{PyBench}, a benchmark encompassing five main categories of real-world tasks, covering more than 10 types of files. 
Given a high-level user query and related files, the LLM Agent needs to reason and execute Python code via a code interpreter for a few turns before making a formal response to fulfill the user's requirements. 
Successfully addressing tasks in PyBench demands a robust understanding of various Python packages, superior reasoning capabilities, and the ability to incorporate feedback from executed code. 
Our evaluations indicate that current open-source LLMs are struggling with these tasks. Hence, we conduct analysis and experiments on four kinds of datasets, proving that comprehensive abilities are needed for PyBench. Our fine-tuned 8B size model: \textbf{PyLlama3} achieves an exciting performance on PyBench which surpasses many 33B and 70B size models.
Our Benchmark, Training Dataset, and Model are available at: \href{https://github.com/Mercury7353/PyBench}{https://github.com/Mercury7353/PyBench}
\end{abstract}

%% file: introduction.tex
\section{Introduction}

\begin{figure}
    \centering
    \includegraphics[width=1\linewidth]{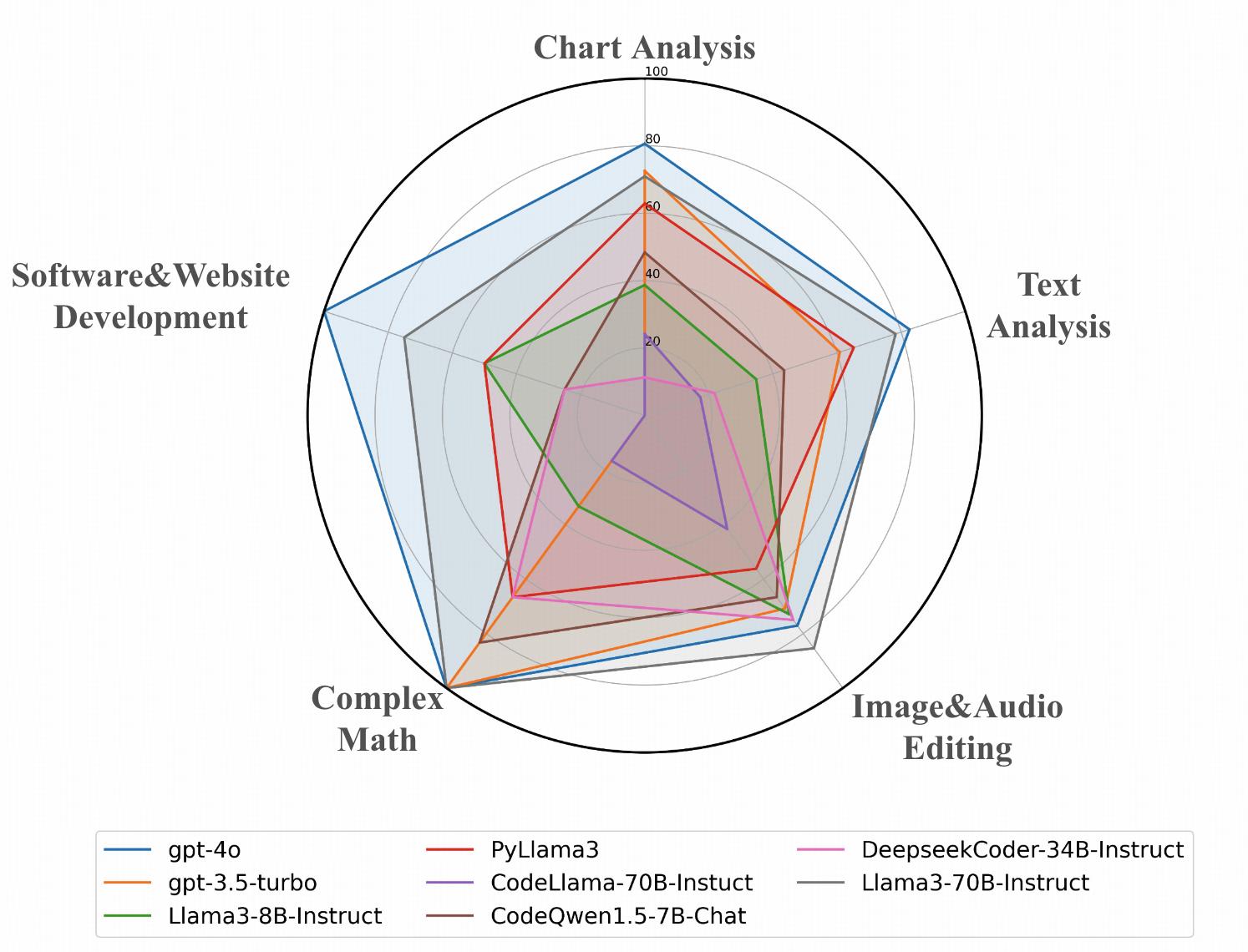}
    \caption{An Overview of LLMs' performance on \textit{PyBench}}
    \label{fig:ini-label}
\end{figure}

\textit{The best tool is the one that gets the job done.} 
Enormous real-world tasks like data analysis and image \& audio processing can be solved by code.
Among many programming languages, Python stands out for its simplicity, ease of use, extensive libraries, and high compatibility, making it a widely used tool for daily tasks.
However, individuals often need to invest significant time in learning how to use extension packages, even if they are already highly proficient in Python itself. 
Thanks to LLM's powerful code capabilities, it can act as an automatic agent \cite{wang2024survey,park2023generative,qin2023toolllm} writing and executing code to solve a wide spectrum of real-world tasks. 

However, current LLM code benchmarks have not covered the real-world task area. HumanEval \cite{chen2021evaluating} and MBPP \cite{austin2021program} focus on function complement and simple Python problems, which could not evaluate the usefulness of LLM Agent. 

%
Although benchmarks such as DS-1000 \cite{lai2023ds}, DevBench \cite{li2024devbench}, and SWE-Bench \cite{jimenez2023swe} focus on repository-level coding issues, they assess the ability of LLMs to use and manage specific codebases. 
These tasks are relatively narrow in scope and inherently limited, deviating from practical daily application scenarios and being overly complex for routine use.

To address the lack of benchmarks for real-world coding tasks, we introduce \textit{PyBench}, a comprehensive and versatile benchmark designed to evaluate the practical coding abilities of LLMs.
Specifically, we formulated real-world coding tasks into \textit{5} main categories: Chart Analysis, Text Analysis, Image \& Audio Editing, Complex Math, and Software \& Website Development. 
Each category consists of comprehensive subclasses of tasks, reflecting real-world situations.
Related files are collected for each subclass, and queries are tailored for the content of files.
For the quantitative evaluation of \textit{PyBench} tasks, we create a set of unit tests to verify whether the tasks are solved successfully. We also employ GPT-4 as a judge to evaluate the solutions and calculate the Average Turns as a measure of problem-solving efficiency. 
We evaluate 3 types of models on \textit{PyBench}: Closed-source LLMs, 70B, 33B, and 7B size Open-source LLMs, and Code LLMs specifically tailored for coding tasks.
The evaluation results indicate most LLMs struggle to solve \textit{PyBench} tasks. 

%
%
%
Consequently, we collect and synthesize four datasets: the homologous training dataset, multi-turn code interaction dataset, multi-turn chat dataset, and code-rich corpus for continue pre-training.
We then conduct a series of analyses and experiments to figure out what are necessary abilities to solve the  \textit{PyBench} tasks and how to improve the LLM performance on real-world coding tasks.
The result demonstrates that merely homologous data cannot help the base model adapt to real-world coding tasks. 
%
The multi-turn code interaction dataset significantly enhances comprehensive capabilities. 
Specifically, the multi-turn chat dataset boosts performance in Chart Analysis, while continued pre-training on a code-rich corpus improves Text Analysis.
%
%
Trained on these datasets, our fine-tuned 8B size model PyLlama3 surpasses Llama3-8B-Instruct on \textit{PyBench}.

In summary, our contributions are as follows:
\begin{itemize}
    \item We construct \textbf{PyBench}, the first comprehensive benchmark for evaluating LLM Agents on real-world coding tasks. PyBench includes real-world files and related queries, covering a wide range of daily situations and file types.
    \item We present PyBench tasks that necessitate the agent's comprehensive capabilities. Using only homologous data is inadequate for adapting base models to real-world coding tasks. The agent's ability for multi-turn interaction is crucial, and continued pre-training on a code-rich corpus is also beneficial.
    \item We conduct continued pre-training of Llama3-8B-Base on a code-rich corpus and fine-tune it on homologous, multi-turn code, and multi-turn chat datasets. Our \textbf{PyLlama3} model achieves outstanding performance on PyBench.
\end{itemize}

\begin{table*}[h]
\centering
\begin{tabular}{p{4cm}p{6cm}p{5cm}}
\hline
\textbf{Category} & \textbf{Subclass} & \textbf{Relative Python Packages}\\
\hline
Chart Analysis & Data Preprocessing, Data Visualization, Machine Learning...& pandas, numpy, sklearn, matplotlib... \\
Text Analysis & Text Based QA, Theme Analysis, Wordcloud... & jieba, wordcloud, PyPDF2... \\
Image \& Audio Editing & Image Generation, Sound Feature Extraction... & opencv, PIL, pydub... \\
Complex Math & Large Number Calculation, Calculus...&numpy, scipy...\\
Software \& Website Development & Game Design, Website Design... & pygame, bs4... \\
\hline
\end{tabular}
\caption{Type of tasks and related Python packages of \textit{PyBench}. We only select commonly used Python packages for these tasks, though other Python packages may also be used for specific tasks.}
\label{catgory}
\end{table*}

%% file: related_works.tex
\section{Related Works}
\subsection{Benchmark on coding ability}
Many existing benchmarks focus on LLM's code ability. HumanEval \cite{chen2021evaluating} and MBPP \cite{austin2021program} are two widely recognized benchmarks primarily evaluating LLM's ability to complete functions or solve simple Python problems. HumanEval-X, HumanEval+, and MBPP+ \cite{zheng2024codegeexpretrainedmodelcode,liu2023codegeneratedchatgptreally} extend the benchmarks by adding multilingual and plenty of extra tests. APPS \cite{hendrycksapps2021} focuses on writing code from natural language description. TACO \cite{li2023taco} builds a more complex benchmark evaluating LLM on algorithmic code tasks.
XCodeEval \cite{khan2023xcodeevallargescalemultilingual} and CodeScope \cite{yan2024codescopeexecutionbasedmultilingualmultitask} provide execution-based multilingual code benchmarks but without real interaction with files.  
MINT \cite{wang2023mint} and $M^3$ToolEval \cite{wang2024executable} aim to evaluate models' multi-turn interaction code ability with tools or human feedback. 
There are also extremely complex and hard benchmarks evaluating LLMs on software development \cite{qian2023communicative,hong2023metagpt}, code repository issues \cite{jimenez2023swe,li2024devbench,zhang2023repocoderrepositorylevelcodecompletion,du2023classevalmanuallycraftedbenchmarkevaluating}, and data science tasks\cite{lai2023ds}.
However, these benchmarks are all limited to specific scenarios.
To the best of our knowledge, no existing benchmark evaluates LLM Agent on real-world coding tasks interacting with files in various situations.  

\subsection{Code LLMs}
Previous works enhance LLM's coding ability through various methods. 
OpenCodeInterpreter \cite{zheng2024opencodeinterpreter} introduces the CodeFeedback dataset and a code execution system with feedback. The fine-tuned model achieves a great performance on coding benchmarks. 
CodeAct \cite{wang2024executable} uses executable Python code to unify LLM agents' action space, enabling sophisticated task execution through multi-turn interactions.
NexT \cite{ni2024next} teaches LLM reasoning the execution process of code step by step, effectively improving the code quality.
WizardCoder \cite{luo2023wizardcoderempoweringcodelarge}, Magicoder \cite{wei2024magicoderempoweringcodegeneration}, and AlchemistCoder \cite{song2024alchemistcoder} build effective fine-tuning datasets from massive and multi-source data to train advanced code LLMs.
Pre-training on code-rich data is also a good method to help LLM coding.
CodeQwen \cite{bai2023qwen}, and DeepSeek-Coder \cite{guo2024deepseek,deepseekai2024deepseekcoderv2breakingbarrierclosedsource} develop specialized models for coding by continuing to pre-train on code data and employing supervised fine-tuning strategies.
\subsection{LLM Agent for real-world tasks} 
LLM as Agent is a great \cite{qian2023communicative,park2023generative,chen2024agent} attempt utilizing LLM in real-world tasks.
ReAct \cite{yao2022react} first introduced Agent's Reasoning and Action Format.
Previous works design many frameworks that build and organize LLM Agents to complete real-world coding tasks.
MetaGPT \cite{hong2023metagpt} ChatDev \cite{qian2023communicative}, DataInterpreter \cite{hong2024data}, and MatplotAgent \cite{yang2024matplotagent} employ agents to complete software development or data science tasks.
AgentCoder \cite{huang2023agentcoder} focuses on simple code complement tasks. Furthermore, some general multi-agent systems try to adapt agents to various tasks \cite{chen2023agentverse,chen2023autoagents,wang2023unleashing}.

%% file: bench.tex
\section{PyBench}

\begin{figure*}[h]
    \centering
    \includegraphics[width=1\textwidth]{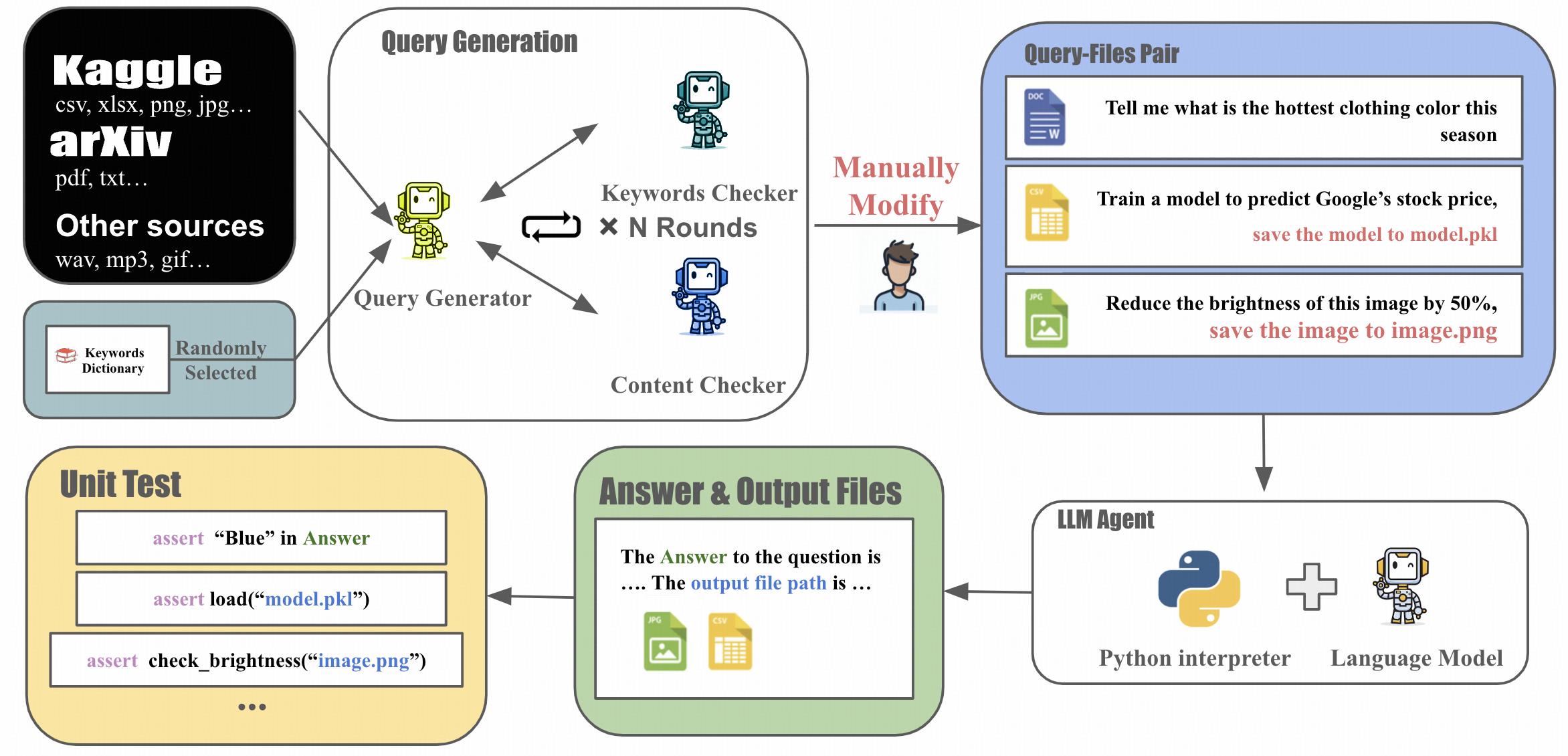} 
\caption{The construction and evaluation workflow of \textit{PyBench}}
\label{Figure 1}
\end{figure*}

Coding is a core skill for LLM Agents. When the agent needs to solve real-world coding tasks, it should not only write executable code but also utilize the results of execution to guide subsequent actions and interact with files.
Python is a powerful programming language with almost 4 million packages, capable of covering almost all real-world coding tasks. 
Therefore, we propose building \textit{PyBench} to evaluate the LLM agent's ability in reasoning, writing executable Python code, and utilizing code results. 

\subsection{Task Formulation} 

We first define what kinds of tasks need to be solved as a truly helpful LLM Agent equipped with a Python code interpreter. Given a user query $q$ described in natural language and the related file $F_{in}$, the Agent need to generate a formal answer $Ans$ to user query and output file $F_{out}$, which fulfill the user's requirement: 

\begin{equation}
Ans, F_{out} = A(q, F_{in})
\end{equation}

\noindent
where $A$ is the LLM Agent equipped with a code interpreter.
To be specific, the query $q$ from users can be very high level, reflecting common occurrences in daily life where users may not have precise requirements.
The related file $F_{in}$ contains various types of data such as chart, text, audio, image, etc., showcase the LLM Agent should adapt to various real-world coding tasks. 

\subsection{Task Categories}

In the realm of practical coding applications, the LLM Agent is required to autonomously tackle a diverse array of real-world coding challenges. 
To ensure a comprehensive evaluation of its capabilities, we have meticulously curated five main categories of real-world coding tasks. 
Each category is designed to test the LLM Agent's proficiency across a broad spectrum of scenarios, ranging from data analysis to software development.
Table \ref{catgory} demonstrates these categories:

\noindent\textbf{Chart Analysis.}
In the digital age, the ability to efficiently analyze and interpret data is indispensable. 
This category focuses on tasks that involve handling \textit{csv} or \textit{xlsx} files for various purposes such as data preprocessing, transformation, visualization, and the application of machine learning algorithms \cite{hong2024data,yang2024matplotagent}. 
The output from the LLM Agent includes detailed analytical reports, visual representations, or even trained machine learning models in \textit{pkl} or \textit{joblib} formats.

\noindent\textbf{Text Analysis.}
This category encompasses tasks related to processing \textit{txt} and \textit{pdf} files, including summarization, keyword extraction, word cloud generation, and thematic analysis.

\noindent\textbf{Image \& Audio Editing.}
As visual content becomes increasingly prevalent, the demand for personalized image-processing solutions rises.
This category evaluates the LLM Agent's ability to manipulate images in \textit{png} and \textit{jpg} formats through various techniques such as saturation adjustment, merging, cropping, and generating QR codes from scratch. 
Parallel to image processing, this category also focuses on the manipulation of audio files (\textit{e.g.} \textit{mp3} and \textit{wav}).
Tasks include volume control, audio trimming, and the creation of audio visualizations, reflecting the diverse needs of users in the realm of audio content creation and modification.

\noindent\textbf{Complex Math.}
Beyond the capabilities of basic calculators or the Chain of Thought methodology \cite{wei2022chain}, this category presents challenges involving large-scale computations, polynomial equation solving, and advanced calculus.
It is designed to test LLM Agents' ability to navigate and solve intricate mathematical problems via a code interpreter.

\noindent\textbf{Website \& Software Development.}
This category is dedicated to the practical application of coding skills in the development of personal websites and simple software projects, such as a Pac-Man game.
It assesses the LLM Agent's ability to translate user requirements into functional and interactive applications, showcasing its versatility and creativity in software development \cite{hong2023metagpt,qian2023communicative}.

\subsection{Data Collection}

We collect and filter the files in \textit{PyBench} from two main sources.

\noindent\textbf{Kaggle Data.}
\href{https://www.kaggle.com/}{Kaggle} is a great platform for machine learning, which contains massive datasets.
We obtained \textit{csv} and \textit{xlsx} data on Kaggle through web crawlers.
There are two principles of filtering the files.
Firstly, the files should not be too large, considering the limited memory in the test environment.
Secondly, the data tables must contain multiple columns with clear meaning, simulating commonly used files.

\noindent\textbf{arXiv Data.}
We collect \textit{pdf} and \textit{txt} from \href{https://arxiv.org/}{arXiv}.
The papers on arXiv are high-quality text with a clear theme and structure, which is suitable for text-based QA, Theme Analysis, and drawing word clouds.   

\noindent\textbf{Other Sources Data.}
For other file types, we responsibly collect files, including \textit{png, jpeg, gif, mp3}, and \textit{wav} from the internet, ensuring that all content respects copyright laws, protects user privacy, and is free from harmful elements.

\subsection{Task Generation}
The queries in \textit{PyBench} must be precisely related to files and diverse to ensure comprehensive coverage. To generate these queries, we designed a multi-agent cooperation mechanism. Figure \ref{Figure 1} illustrates the \textbf{PyBench} construction workflow. 

First, we prepare lists of keywords for each subclass in a category, forming a keywords dictionary for each category \cite{eldan2023tinystories}. The type of file determines which category the task belongs to. Given the initial lines of a file(for chart and text files) and randomly selected keywords from the related category's dictionary, the Query Generator selects appropriate keywords that align with the file content and compose a query. Detailed keyword list examples can be found in Appendix \ref{KeyWordLists}.

Next, a Keywords Checker and a Content Checker verify if the query aligns with the file content and the selected keywords. Feedback from these checkers is used to refine the query. Once both checkers approve, the query undergoes manual editing
to clarify the output format and file path for the convenience of verifying. If the Query Generator fails to produce a proper query within \textit{5} rounds, these files will be skipped. 

Queries that pass all checks are paired with their related files to form a query-file pair, which constitutes a task in \textit{PyBench}.
Appendix \ref{QueryGenerationPrompt} and \ref{CheckerPrompt} detailed the prompt of each agent.

\subsection{Evaluation}
\subsubsection{Trajectory Generation}
\label{TrajGen}

\noindent\textbf{Equip LLM Agent with Code Interpreter.}
We equipped each LLM with a code interpreter to execute Python code written by the LLM and provide feedback on the execution results.
Previous works on LLM tool usage \cite{qin2023toolllm} typically used tools in a function-calling format.
Inspired by \cite{wang2024executable}, which uses code as an action and outperforms alternatives, we designed two special tokens: \textit{<|execute\_start|>} and \textit{<|execute\_end|>} to help the LLM use the code interpreter more effectively.
Appendix \ref{CodeVsFunction} shows the different performance of the two formats.

\begin{figure}[ht]
\centering 
\includegraphics[width=0.5\textwidth]{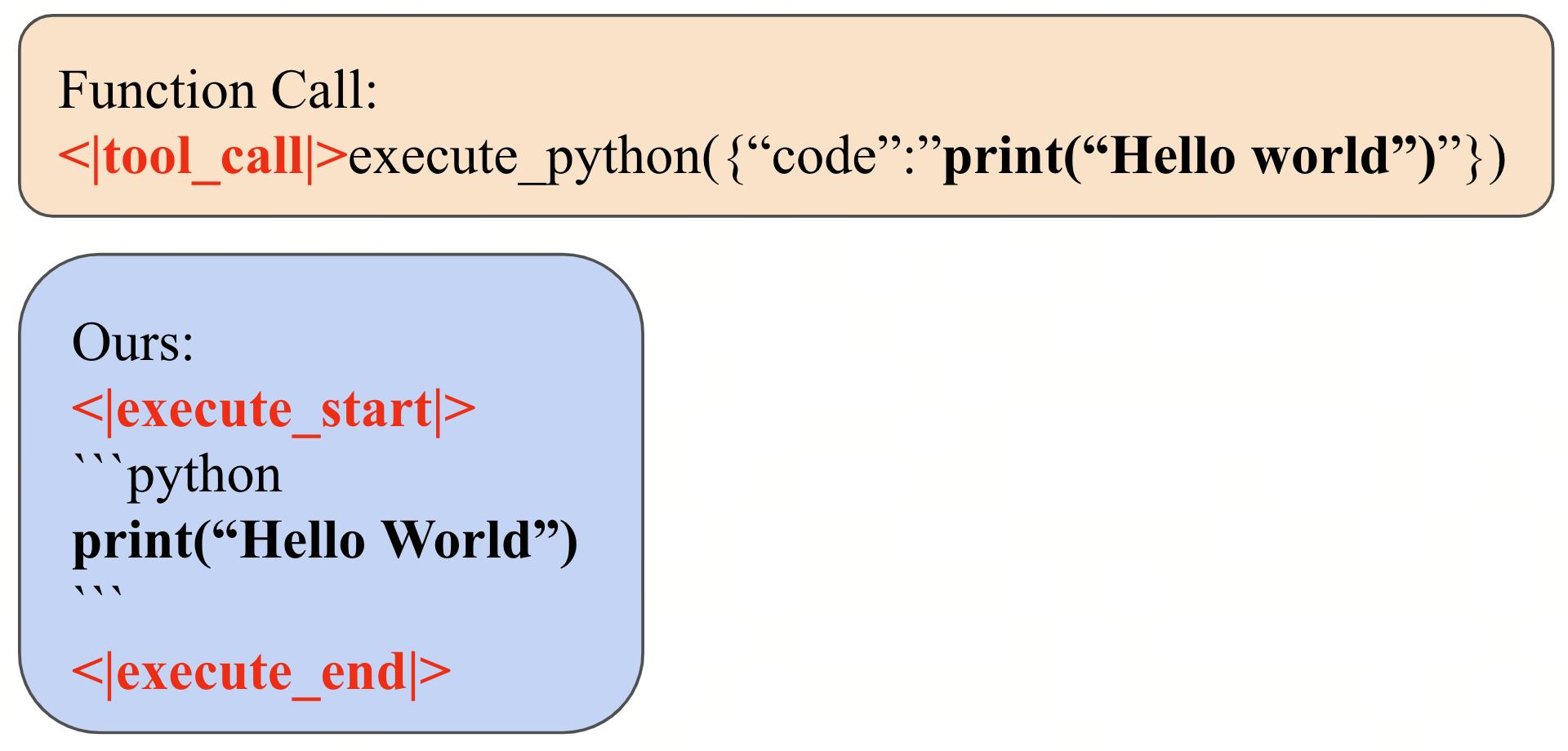} 
\caption{Function Call vs. Our Code Interpreter Format} 
\label{fig:2}
\end{figure}

\begin{figure}[ht]
\centering 
\includegraphics[width=0.5\textwidth]{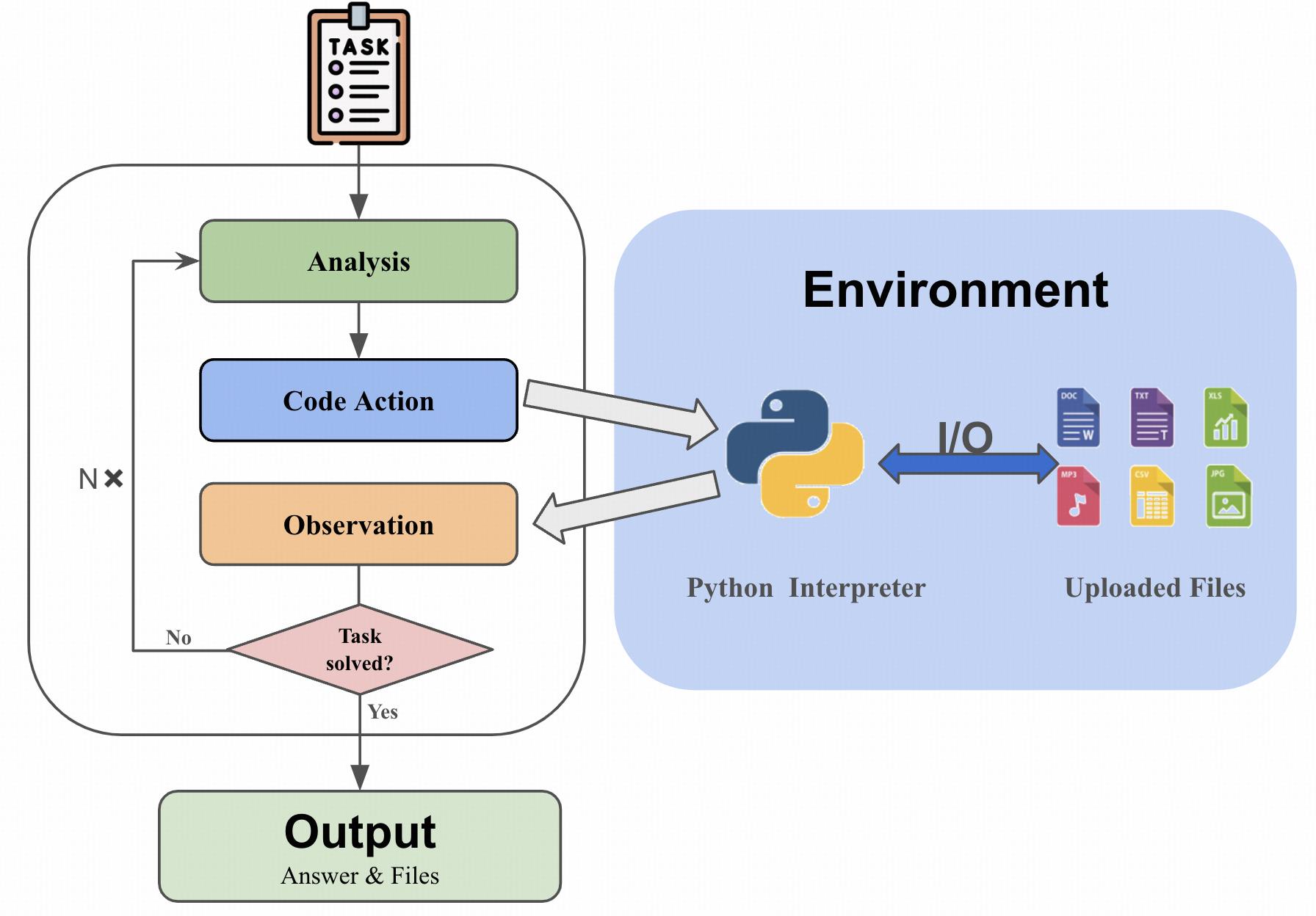} 
\caption{Generating Trajectory Data by ReAct} 
\label{fig3}
\end{figure}

\noindent\textbf{Reasoning and Action.}
Given the task description and the uploaded file path, the LLM Agent is prompted to analyze the current situation and plan its actions before writing executable code \cite{yao2022react}.
The code will be executed in a pre-defined Python environment (with commonly used packages).
The result of the code, whether is an output or an error message, well serve as feedback to the LLM.
The LLM will follow the loop until it fulfills the task or reaches a maximum step limit (default set to 10).
Figure \ref{fig3} detailed this process.

\subsubsection{Unit Test}
To objectively and effectively test whether the LLM Agent has completed a task, we have implemented a unit test for each task in \textit{PyBench}.
For tasks with a fixed answer, we verify whether the Agent provides a final response that contains the correct answer.
For tasks requiring file output, such as cleaning datasets or editing images and audio, we check whether the output files meet the specified requirements.
For tasks without a fixed answer, such as generating a word cloud or a website, we verify the existence of the output files. The design of the unit Test is detailed in Appendix \ref{UnitTest}

\subsubsection{LLM as Evaluator}
Although unit tests are convenient and objective, they may fail to comprehensively evaluate open-ended tasks, such as assessing the coherence and fluency of text output by large models.
Therefore, we also employ an LLM (GPT-4o) as an evaluator to provide pass-or-fail decisions for each trajectory, serving as an alternative to unit tests. Appendix \ref{evaluationPrompt} provides the detailed prompt used for the LLM evaluator.
\begin{table*}[h]
\centering
\begin{tabular}{p{4cm}p{2cm}p{1cm}p{1cm}p{1cm}p{1cm}p{3cm}}
\hline
\textbf{Datasets} &   \textbf{Instructions} & \multicolumn{4}{c}{\textbf{Turns per Traj.}} & \textbf{Tokens per Traj.} \\
\cline{3-6}
   &  & \textbf{<=3} & \textbf{4to6} & \textbf{7to9} & \textbf{>=10} &  \\
\hline
CodeFeedback  & 66383 & 41696 &  21583& 2923 &  181  &1503.93  \\
CodeActInstruct  & 7139 & 3482 & 3567 &0  & 0   &1165.03  \\
PyInstruct  & 3091 &1234  & 1644 & 164   & 49 & 2017.15 \\
\hline
\end{tabular}
\caption{Statistics of \textit{CodeFeedback}, \textit{CodaActInstruct}, and \textit{PyInstruct}. Token statistics are computed by using Llama-2 tokenizer.}
\label{statistics}
\end{table*}

\subsubsection{Evaluation Metrics}
There are three evaluation metrics in \textit{PyBench}.

\noindent\textbf{Pass Rate (UT).} The percentage of passed tasks evaluated by unit tests. 

\noindent\textbf{Pass Rate (LLM).} The percentage of passed tasks evaluated by the LLM Evaluator. 

\noindent\textbf{Average Turns.} The number of steps taken to complete each task. If a task fails, the number of turns is set to the maximum turn limit (10 turns).






%% file: method.tex
\section{Fine-tuning LLM Agent for Real World Coding Task}

In order to figure out what kind of capabilities are required and what training data could help enhance the LLM Agent's performance on \textit{PyBench}, we collect 4 datasets enhancing the Agent's abilities in planning, coding, and multi-turn interaction.

\noindent\textbf{Homologous dataset.}
%
%
%
%
Intuitively, homologous datasets can enhance performance on similar tasks. Therefore, we employ GPT-3.5-turbo to synthesize a homologous dataset: \textit{PyInstruct}.
We generate tasks using the same method as PyBench, but with different files and without manual modifications. Subsequently, using the method described in Section \ref{TrajGen}, we synthesized 3,091 trajectories based on these tasks. These trajectories cover every task category in PyBench, resulting in the construction of \textit{PyInstruct} as a homologous dataset to PyBench.
\noindent\textbf{Multi-turn of code interaction dataset.}
Most of the tasks in \textit{PyBench} need multi-turn interaction with the Python code interpreter. 
It will provide feedback on code execution results or error traceback messages, which should be fully leveraged by the LLM Agent. 
%
There are several existing datasets aiming to enhance LLM's ability.
\textit{CodeActInstruct} \cite{wang2024executable} focuses on improving LLM's abilities in various multi-turn tasks such as information seeking, software tool usage, external memory access, and robot planning, all executed in the format of Python code.
\textit{CodeFeedback} \cite{zheng2024opencodeinterpreter} filters open-source code instruction data and converts them into multi-turn code with execution results.
We repurpose the data by ``equipping'' our special token for the Python code interpreter. 
These datasets may help enhance LLM's ability to utilize code feedback. 
Table \ref{statistics} shows the statistics information of  \textit{CodeFeedback}, \textit{CodeActInstruct}, and \textit{PyInstruct}.

\noindent\textbf{Multi-turn chat dataset.}
Additionally, in \textit{PyBench}, the LLM Agent is expected to comprehend the user's instructions and provide a formal response upon completing the task.
High-quality multi-turn chat is also crucial for a Code Agent.
\textit{UltraChat} \cite{ding2023enhancing} is a large-scale dataset comprising 1.5 million high-quality multi-turn instructional dialogues specifically designed to improve the performance of open-source conversational models, which is perfectly aligned with our requirements.

\noindent\textbf{Code-Rich corpus.}
The foundation ability of a code agent is the quality and correctness of its code.
We assume that continue pre-train on code-rich corpus could contribute to solving \textit{PyBench} tasks. 
\textit{The-stack-v2} \cite{lozhkov2024starcoder} introduces a code-rich corpus of Jupyter notebooks, which contains 11 million lines. 

After collecting the four types of datasets, we conducted a series of experiments to figure out what are necessary abilities to solve the \textit{PyBench} tasks and how to improve the LLM performance on real-world coding tasks.

%% file: experiment.tex
\section{Experiment}

\begin{table*}[!h]
\small 
\setlength{\tabcolsep}{1pt} 
\renewcommand{\arraystretch}{1.5}
\begin{tabular}{p{3cm}p{1cm}p{0.8cm}p{1cm}p{1cm}p{1cm}p{1cm}p{1.5cm}p{0.8cm}p{0.8cm}p{2.0cm}p{1.3cm}}
\cline{1-12}
\textbf{Models} & \textbf{Params.} & \multicolumn{8}{c}{\textbf{PyBench}} & \textbf{HumanEval(+)} & \textbf{MBPP(+)} \\
\cline{1-12}
   &  & \textbf{Pass Rate(LLM)} & \multicolumn{6}{c}{\textbf{Pass Rate(UT)}} & \textbf{Avg Turns} \\
\cline{4-9}
   &  &  &  {\textbf{Chart.}} &  {\textbf{Text.}} &  {\textbf{Image.}} &  {\textbf{Math.}} &  {\textbf{Software.}} &  {\textbf{Overall}} \\
\cline{1-12} 
\textit{Closed-source Models} \\
\cline{1-12} 
GPT-3.5 Turbo & - & 58.3 & 72.6 & 60.9 & 70.8 & 100.0 & 0.0 & 62.9 & 5.2 &78.6(70.7)&82.5(69.7) \\
GPT-4o & - & \textbf{81.8} & \textbf{80.6 }& \textbf{82.6} & 77.1 & 100.0 & 100.0 & 79.7 &\textbf{4.3}&\textbf{91.5}(85.4)&\textbf{91.2}(76.6) \\
GPT-4o-mini & - &  67.1 & 77.4 & 69.6 & \textbf{87.5} & 100.0 & 100.0 & \textbf{81.1} & 4.5 &87.2(85.4)&90.5(76.5)\\
\cline{1-12}  
\textit{70B size} \\
\cline{1-12} 
Llama3-Instruct & 70B & \textbf{80.2} & \textbf{38.7} & \textbf{34.8} & \textbf{72.9} & 33.3 & 50.0 & \textbf{78.3} & \textbf{4.0}&\textbf{77.4(72.0)}&\textbf{82.3(69.0)} \\
CodeLlama-Instruct & 70B & 26.6 & 24.2 & 17.4 & 41.7 & 16.7 & 0.0 & 37.8 & 7.5&72.0(65.9)&62.2(51.2) \\
DeepSeek-llm-chat & 67B & 39.4 & 21.0 & 17.4&37.5&33.3&25.0&26.6&9.2&67.7(58.5) & 62.2(50.5)\\
\cline{1-12} 
\textit{33B size} \\
\cline{1-12} 
DeepSeek-coder-Instruct & 33B & 35.3 & 22.6 & 0.0 & 4.2 & 16.7 & 0.0 & 37.0 & 6.7&81.1(75.0)&80.4(70.1) \\
CodeLlama-Instruct & 34B & 9.6 & 8.1 & 13.0 & 0.0 & 16.7 & 0.0 & 6.3 & 8.9&51.8(43.9)& 69.3(56.3) \\
Yi-1.5-Chat-16K & 34B &44.2&33.9&34.8&56.3&50.0&0.0&41.3 & 9.0 & 58.8(53.4) & 77.5(63.6) \\
Qwen-1.5-Chat & 32B & 43.0 & 12.9 &39.1&10.4&50.0&75.0&19.6&9.3& 58.5(55.5) & 66.1(56.1) \\
Gemma-2-Instruct & 27B & \textbf{66.1} & \textbf{58.1}&\textbf{56.5}&\textbf{87.5}&\textbf{83.3}&75.0&\textbf{69.2}&\textbf{5.0}&49.4(45.7)&73.3(61.1)\\
\cline{1-12} 
\textit{7B size} \\
\cline{1-12} 
Llama3-Instruct & 8B & 49.7 & 43.5 & 26.1 & 72.9 & 66.7 & 50.0 & 49.7 & 6.7&61.6(56.7)&70.1(59.3) \\
Mistral-Instruct-v0.2& 7B & 17.5&17.7&17.4&18.8&16.7&0.0&17.5&8.9&35.4(30.5)&31.0(24.1)\\
Gemma-2-Instruct & 9B & 41.7&37.1&30.4&70.1&83.3&50.0&49.7&7.2&55.5(48.8)&66.1(56.1) \\
Qwen2-Instruct & 7B & 57.8 & 59.7 & 47.8 & 58.3 & 83.3 & \textbf{75.0} & 58.7 & \textbf{5.9}&64.6(61.0)&62.3(52.3) \\
CodeQwen1.5-chat & 7B & 49.2 & 48.4 & 43.5 & 66.7 & 83.3 & 25.0 & 54.5 & 7.9&\textbf{83.5(78.7)}&\textbf{79.4(69.0)} \\
CodeActAgent-Llama2 & 7B & 12.4 & 16.1 & 21.7 & 20.8 & 33.3 & 0.0 & 18.9 & 8.1&18.9(15.2)&22.8(18.3) \\
CodeActAgent-Mistral & 7B & 18.8 & 17.7 & 4.3 & 20.8 & 16.7 & 0.0 & 16.1 & 8.8&28.7(26.8)&43.9(35.4) \\
OpenCodeInterpreter-DS & 6.7B & 25.0 & 11.3 & 21.7 & \textbf{75.0} & 66.7 & 25.0 & 51.7 & 6.8&77.4(73.8)&80.2(66.4) \\
InternLM2\_5-chat & 7B &17.5&32.3&13.0&4.17&50.0&25.0&20.27&9.84& 57.3(51.2) & 61.6(51.5) \\
\rowcolor{green!20}
\textbf{PyLlama3(w/o cpt)} & 8B & 56.7 & 58.0 & 52.2 & 73.0 & 50.0 & 50.0 & 58.7& 6.3&54.3(48.2)&59.5(50.3) \\
\rowcolor{green!20}
\textbf{PyLlama3} & 8B & \textbf{73.4} & \textbf{62.9} & \textbf{65.2} & 56.3 & 66.7 & 50.0 & \textbf{60.8} & 6.1&57.3(48.2)&62.2(51.1) \\
\cline{1-12} 
\end{tabular}
\caption{The main result table. We test closed-source models, 70B size models, 33B size models, and 7B size models. We bold the best results for each size model's results.}
\label{mainresult}
\end{table*}

\begin{table*}[h]
\centering
\renewcommand{\arraystretch}{1.5} 
\setlength{\tabcolsep}{3pt} 
\begin{tabular}{llllllllll}
\toprule
\textbf{PyInst.} & \textbf{Code.} & \textbf{UltraCh.} & \textbf{Jupyter.}&\multicolumn{6}{c}{\textbf{Pass Rate(UT)}}   \\
\cline{5-10}
   &  &  & &\textbf{Chart.} & \textbf{Text.} & \textbf{Image.} &\textbf{Math.} &\textbf{Software.}&\textbf{Over All} \\

\midrule
 \cmark & \xmark & \xmark & \xmark&3.2&13.0&6.3&50.0&25.0&8.4  \\
\xmark & \cmark & \xmark & \xmark &40.3&26.1&58.3&83.3&25.0&45.5 \\
 \cmark & \cmark & \xmark & \xmark&51.6&56.5&60.4&\textbf{100.0}&25.0&56.6  \\
 \cmark & \cmark & \cmark & \xmark&\textbf{62.9}&52.2&56.3&83.3&25.0&58.7  \\
 \cmark & \cmark & \xmark & \cmark&58.0&52.2&\textbf{73.0}&50.0&\textbf{50.0}&59.4  \\
\rowcolor{green!20}
 \cmark & \cmark & \cmark & \cmark&\textbf{62.9}&\textbf{65.2}&56.3&66.7&\textbf{50.0}&\textbf{60.8} \\
\bottomrule
\end{tabular}
\caption{Pass Rate (UT) with different training datasets. \textit{PyInst.} stands for \textit{PyInstruct}. \textit{Code.} refers to \textit{Codefeedback \& CodeActInstruct}. \textit{UltraCh.} means \textit{UltraChat}, and \textit{Jupyter.} indicates continuing pre-training on the Jupyter notebook corpus.}
\label{DataAblation}
\end{table*}

\subsection{Main Result on PyBench} 

\noindent\textbf{Testing Evaluation Setup.}
Firstly, we prepare a conda environment equipped with 182 commonly used Python packages (Detailed in Appendix \ref{PythonPackages}).
The max turn is set to $k=10$.
We prompt the LLM to use code interpreter and follow ReAct \cite{yao2022react} format. (Appendix \ref{evaluationPrompt})
The code in LLM's response will be extracted, and the execution result will be returned to LLM.
After getting the trajectory and output files, we calculate the pass rate through the unit test set (\textit{UT}) and LLM Evaluator set (LLM).
We also adopt other benchmarks to test the model's basic code ability, including HumanEval \cite{chen2021evaluating} and HumanEval+ \cite{liu2023codegeneratedchatgptreally} for single-turn code generation, and MBPP \cite{austin2021program} and MBPP+ \cite{liu2023codegeneratedchatgptreally} for simple Python programming problems.

\noindent\textbf{Results on \textit{PyBench}.}
Table \ref{mainresult} shows our main experiment result.
GPT-4o scored highly in all five tasks, showing its great ability. 
However, models including CodeLlama-Instruct-70B, DeepSeek-coder-Instruct-33B, and CodeLlama-Instruct-33B, although trained on code corpus and perform well on HumanEval and MBPP, struggle to solve \textit{PyBench} tasks, proving that code ability does not always lead to a good score on \textit{PyBench}. 
Most of the advanced 7B size models can only solve about 50\% of tasks in \textit{PyBench}. 
Our \textit{PyLlama3}, with continued pre-train and fine-tuning on related datasets, surpasses Llama3-8B-Instruct on Chart Analysis, Text Analysis, and Complex Math.


\noindent\textbf{Coding ability is the foundation.}
CodeActAgent-Llama-2-7B and CodeActAgent-Mistral-7B \cite{wang2024executable}, although trained on CodeActInstruct, who teach LLM Agent use tools in code format, have the weakest performance on \textit{PyBench}.
The failure might be caused by low performance on basic coding ability benchmarks: HumanEval(+) and MBPP(+). The result illustrates basic code ability and the foundation to solve real-world coding tasks in PyBench.

\noindent\textbf{PyBench evaluates comprehensive abilities of LLM Agent.} 
Great performance on basic coding benchmarks does not directly lead to equal performance on PyBench. Although models like DeepSeek-coder-Instruct-33B and CodeLlama-Instruct-34B scored well on HumanEval(+) and MBPP(+), they got a weak score on PyBench. This indicates that mere coding ability cannot lead to success in real-world coding tasks. PyBench also evaluates the LLM agent's abilities, including planning, multi-turn interaction, leveraging code feedback, and generating formal responses to the user.

\noindent\textbf{Compare LLM Evaluator and Unit Test.}
Table \ref{mainresult} shows that the evaluation results from the LLM and the unit tests are generally consistent, with only minor fluctuations. However, there are significant discrepancies for several models. To investigate this, we randomly selected 60 trajectories from the entire dataset for manual assessment. We found that in over 80\% of the cases, the LLM evaluation results were consistent with the unit test results. Nevertheless, in some instances, the LLM evaluator might only determine whether the code executes successfully without verifying its correctness. In other cases, the tested LLM agent might write code or provide answers that coincidentally pass the unit tests. A detailed analysis of these scenarios is provided in Appendix \ref{LLMVSUT}.


\subsection{Analysis on the training dataset}

In this section, we conduct ablation studies on training datasets to explore the necessary capabilities for solving \textit{PyBench}.

\subsubsection{Train Setup}
We conduct full-parameter supervised fine-tuning on Llama3-8B with a sequence length of 32768 tokens \cite{liu2023scaling,roziere2023code,touvron2023llama}, ensuring it could handle the content of chart and text files. 
For each version of the supervised fine-tuning model, we use 32 A100 GPU and train 4000 steps.
The learning rate is set as 1e-5 with a 0.05 warm-up ratio and a cosine scheduler.
As for the continue pre-trained model, we added an extra 3000 steps to train the corpus.

\subsubsection{Ablation study}

LLMs' performance on specific tasks can often be enhanced through supervised fine-tuning on homologous datasets, where the model learns exactly what to do in particular situations.
Initially, we hypothesized that training the Llama3-8B-base model on \textit{PyInstruct}, 3k homologous trajectories data, would teach the model to handle real-world coding tasks.
However, as shown in Table \ref{DataAblation}, this model struggles with \textit{PyBench} tasks.
From the generated trajectories, we observed that the model fails to follow human instructions and even struggles to locate the correct file paths.

To address the drawbacks, we add two multi-turn code interaction datasets: \textit{CodeActInstruct} \cite{wang2024executable} and \textit{CodeFeedback} \cite{zheng2024opencodeinterpreter}.
We train the model on these datasets and PyInstruct. The result indicated a significant improvement. But when we remove \textit{PyInstruct} and train the model only on \textit{CodeActInstruct} and \textit{CodeFeedback}, the model's performance suffers a sharp decline in Chart Analysis and Text Analysis, demonstrating \textit{PyInstruct} can enhance the model's capability.

Additionally, to further enhance the model's capabilities, we added \textit{UltraChat} \cite{ding2023enhancing} in the training datasets and trained the model with the same training settings. However, the result shows adding \textit{UltraChat} only brings improvement to Chart Analysis tasks.

We found that the model still struggles with using some packages and specific techniques for real-world coding tasks. 
Consequently, we conducted continue pre-train on a Jupyter Notebook corpus \cite{lozhkov2024starcoder} before fine-tuning on \textit{PyInstruct, CodeFeedback, CodeActInstruct}, and \textit{UltraChat}.
As shown in Table \ref{DataAblation}, continue pre-train can also bring great improvement to Text Analysis tasks.
The continued pretraining model achieves the best performance after fine-tuning on \textit{PyInstruct, CodeFeedback, CodeActInstruct}, and \textit{UltraChat}, which we named to \textbf{PyLlama3}.
And the version without continue pre-train is named \textbf{PyLlama3(w/o cpt)}.


In conclusion, we figured out that PyBench requires the model's basic coding abilities, which instruct them to write Python properly, as well as multi-turn interaction and reasoning abilities to solve real-world tasks.


%% file: conclusion.tex
\section{Conclusion}
\label{sec:bibtex}

In this paper, we propose \textbf{PyBench}, a comprehensive benchmark encompassing five main categories that reflect real-world coding situations.
Through collecting files and generating related queries, PyBench could evaluate LLM's usability and efficiency in real-world tasks. 
After evaluating plenty of LLMs, we find many of them are struggling with real-world coding tasks. 
We collected and synthesized four datasets and trained \textbf{PyLlama3} whose performance surpass many 7B, 33B, and 70B size models. 
Our ablation studies prove the effectiveness of the datasets, figuring out a way to train a model that could adapt to real-world coding tasks. 
Solving PyBench tasks means the LLM could interact with the file system with Python code, which symbolizes a great milestone in developing a really usable LLM Agent who can serve as a helpful life assistant for humankind.


%% file: limitations.tex
\section{limitations}
Our work introduces a comprehensive benchmark: PyBench to evaluate LLM Agent on real-world coding tasks. Although five main categories are included in PyBench, there are many cases in the real world that have not been covered. The coding problem that uses other coding languages instead of Python is not covered either.

%% file: appendix.tex
\appendix

\newpage

\section{Prompts}

\subsection{Equip LLM with a code interpreter}  
\begin{figure}[H]
    \centering
    \includegraphics[width=1\linewidth]{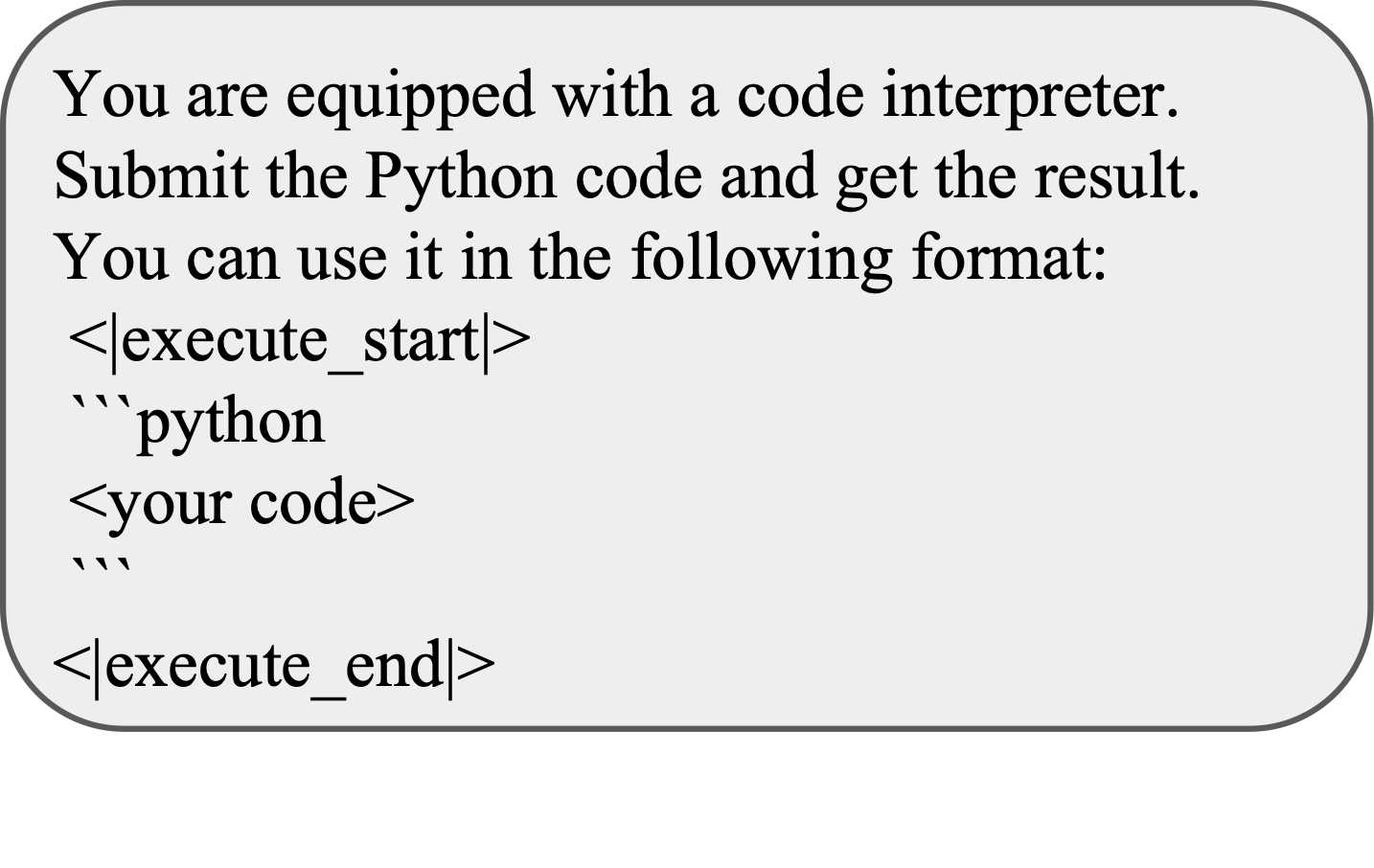}
    \caption{Prompt equipping LLM Agent with a code interpreter}
    \label{codeprompt}
\end{figure}

Fig \ref{codeprompt} shows how we prompt the LLM to use code interpreter.

\subsection{Query Generation Prompt}
\label{QueryGenerationPrompt}
\begin{figure*}
    \centering
    \includegraphics[width=1\linewidth]{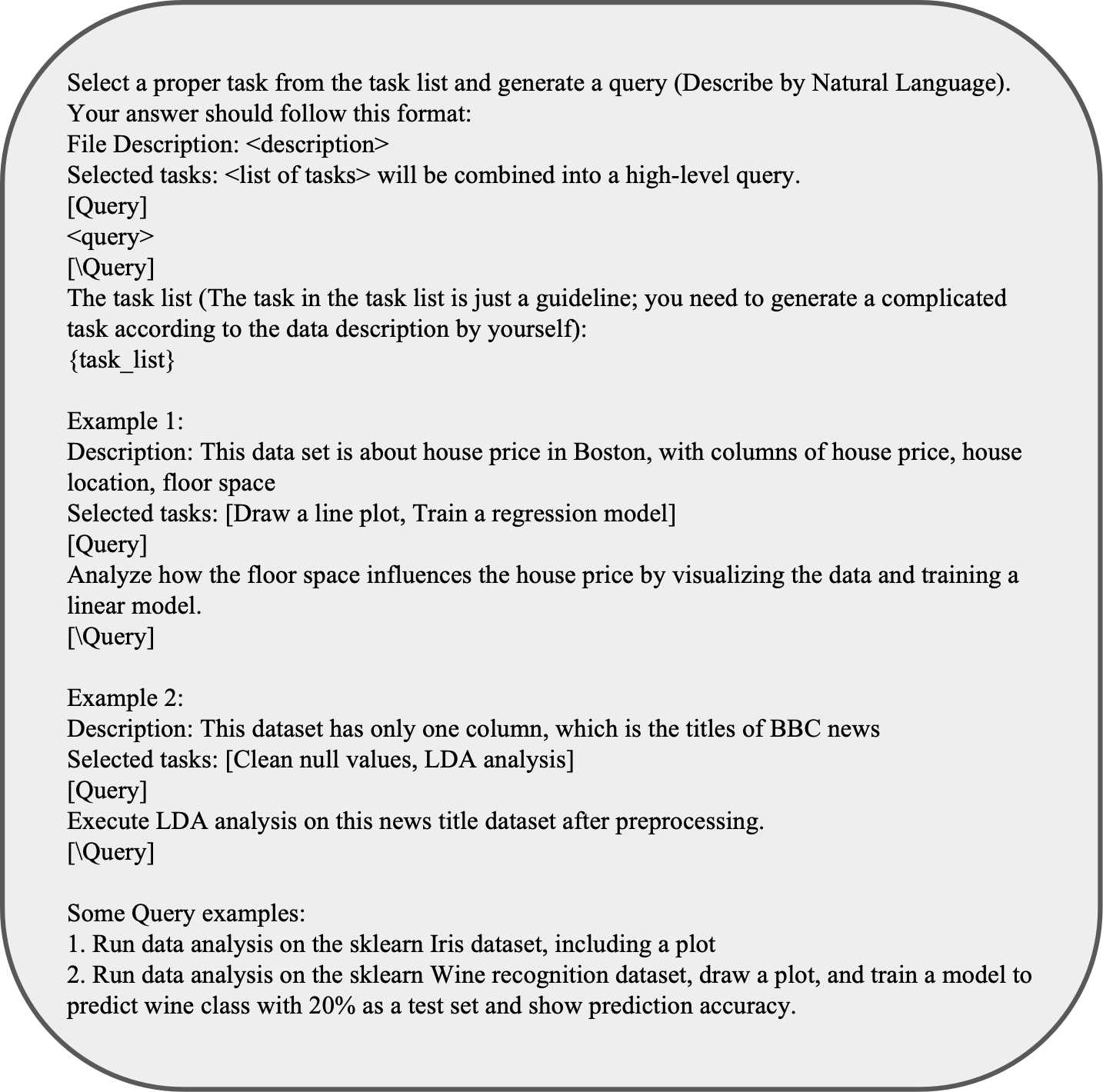}
    \caption{Prompt guide the LLM to generate queries}
    \label{querygen}
\end{figure*}

Fig \ref{querygen} is the prompt to generate a related and diverse query from given file content.




\subsection{Checker Prompt}
\label{CheckerPrompt}
\begin{figure*}
    \centering
    \includegraphics[width=1\linewidth]{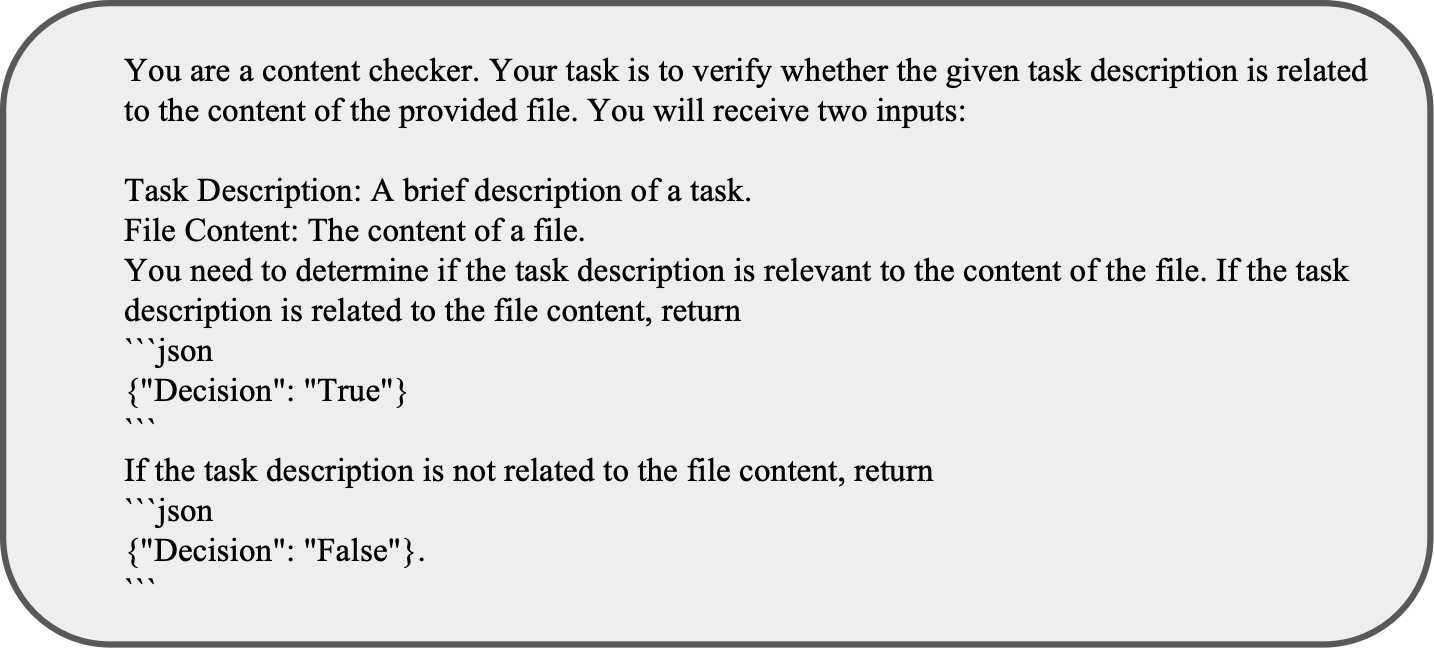}
    \caption{Prompt of The Content Checker}
    \label{checker1}
\end{figure*}

\begin{figure*}
    \centering
    \includegraphics[width=1\linewidth]{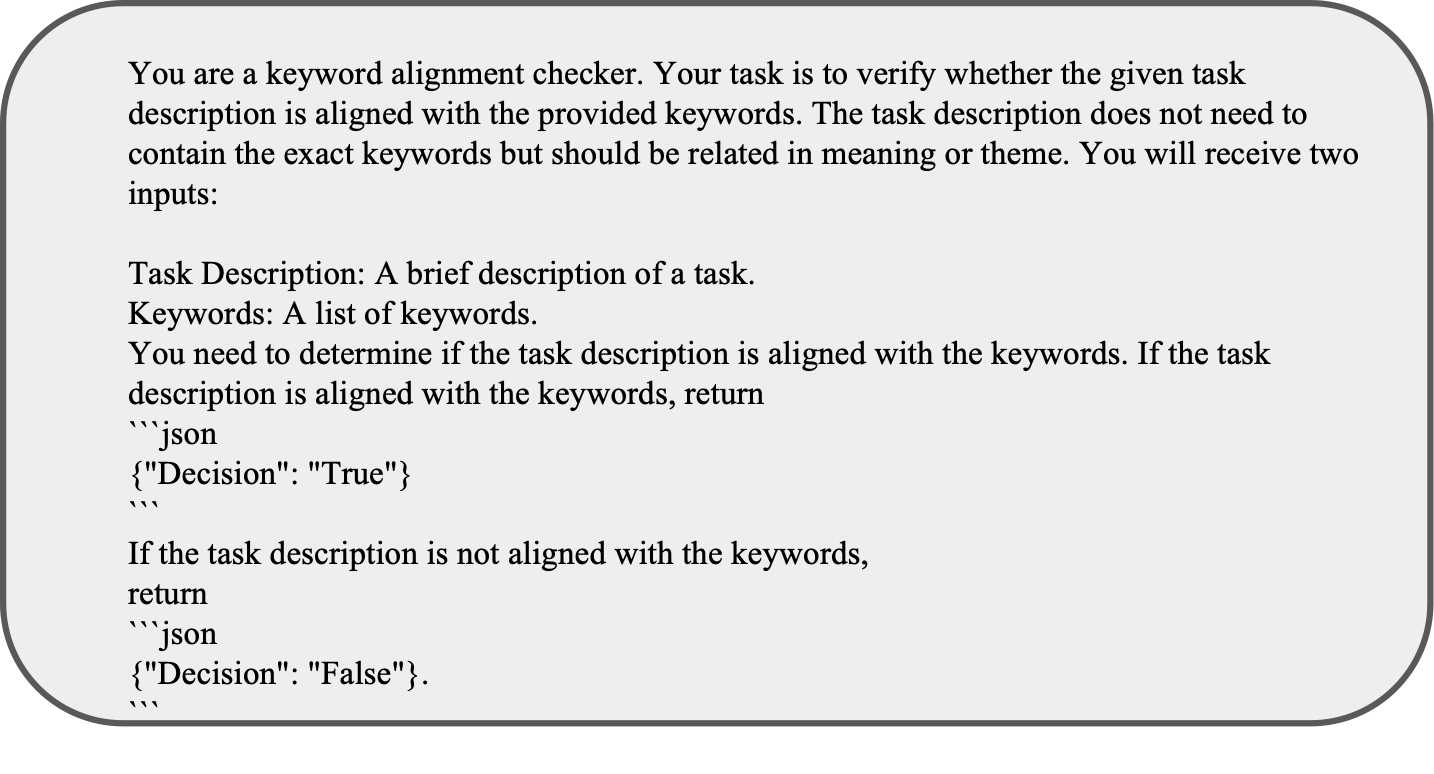}
    \caption{Prompt of the Keyword Checker}
    \label{check2}
\end{figure*}

Fig \ref{checker1} and Fig \ref{check2} are the prompts for the content checker and keywords checker.





\subsection{Evaluation Prompt}
\label{evaluationPrompt}
\begin{figure*}
    \centering
    \includegraphics[width=1\linewidth]{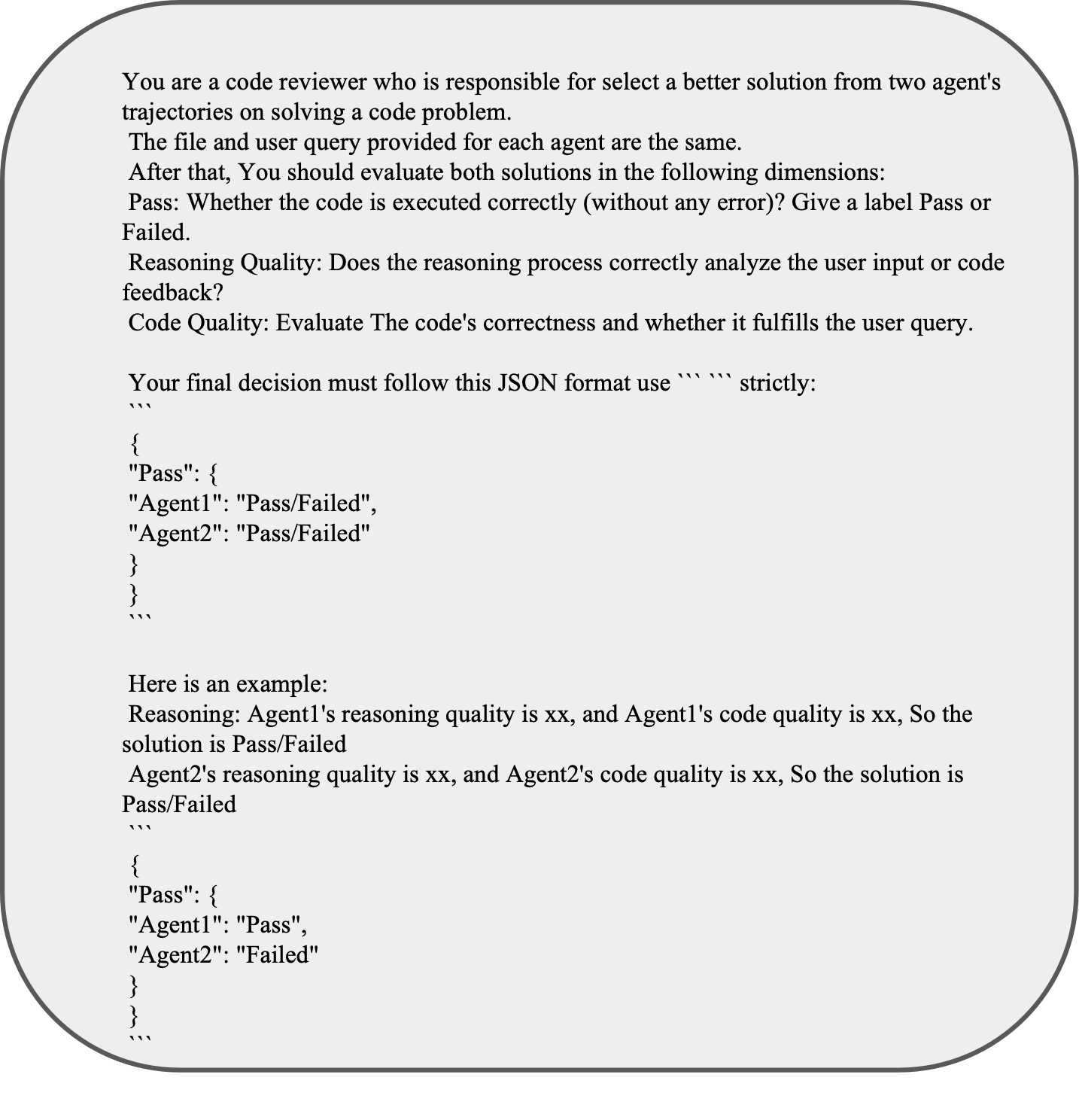}
    \caption{Evaluation Prompt}
    \label{evaPro}
\end{figure*}

Figure \ref{evaPro} shows the LLM Evaluator's prompt.

\section{Key words example for each category}
\label{KeyWordLists}
\begin{figure}
    \centering
    \includegraphics[width=1\linewidth]{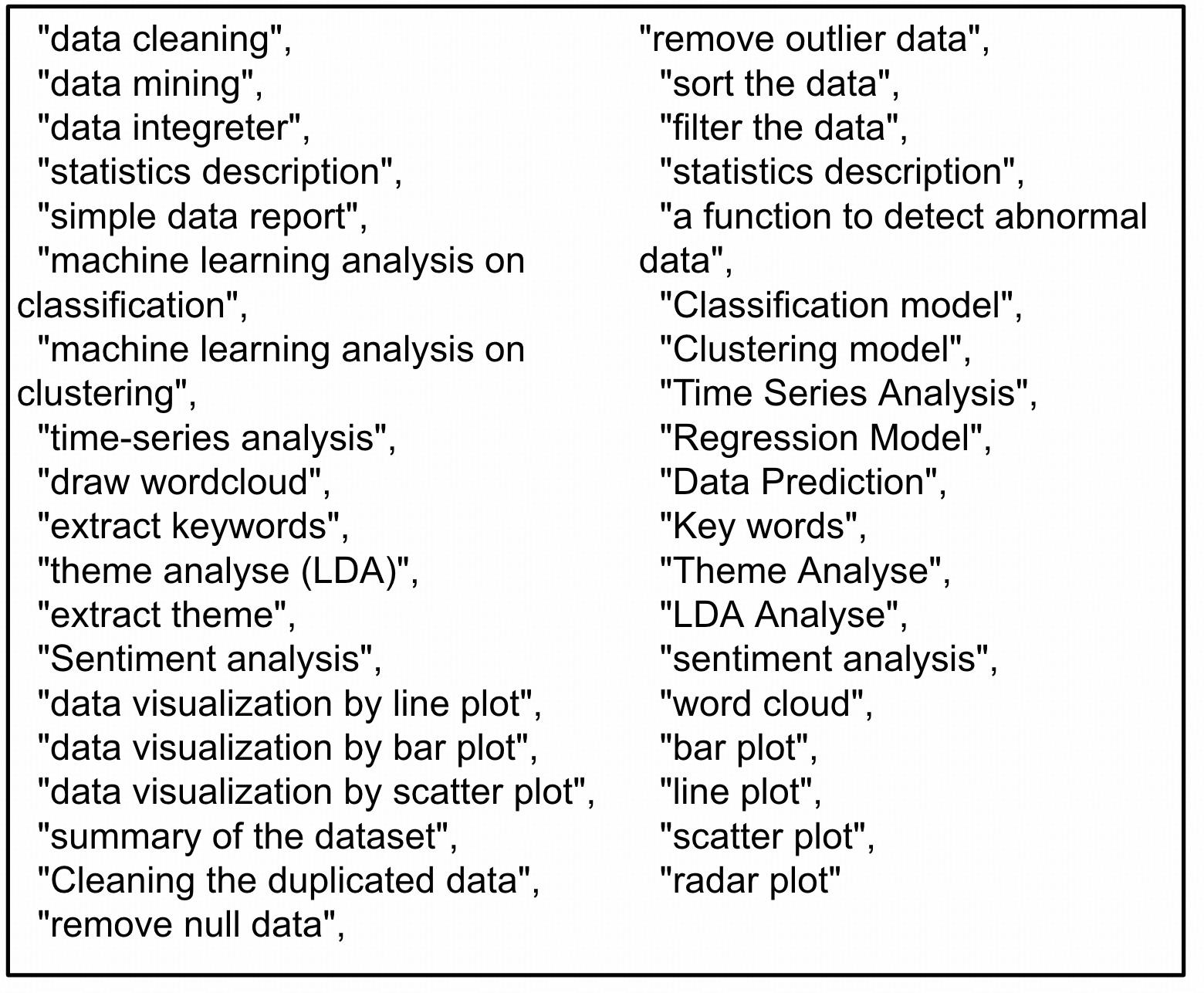}
    \caption{Keywords list on Chart Data Analysis}
    \label{key}
\end{figure}

Fig \ref{key} is a keywords list example for Chart Analysis tasks

\section{Example Unit Test}
\label{UnitTest}
We show three types of Unit Tests in this appendix.
\subsection{Directly Verify the Answer}
For the task with a fixed answer, we check whether the final response contains the answer
\begin{lstlisting}[language=Python]
def test_task_19(trajectory):    
    final_answer=trajectory[-1]['content']
    answer_list=["-346","2504"]
    for ans in answer_list:
        assert ans in final_answer
\end{lstlisting}
\subsection{Verify the output file}
For task clarify the output file, we check whether the output file fulfills the requirements.
\begin{lstlisting}[language=Python]
def test_task_88(trajectory):
    image_path = "./output/88.png"
    ref_path = "./data/88.jpeg"
    image = Image.open(image_path)
    gray_image = image.convert('L')
    gray_array = np.array(gray_image)
    black_threshold = 50
    black_pixels = np.sum(gray_array <= black_threshold)
    total_pixels = gray_array.size
    black_ratio = black_pixels / total_pixels
    assert black_ratio > 0.25, "The black pixel percentage does not exceed 30%."
\end{lstlisting}
\subsection{Verify the output file path}
For Open-end tasks like website development, we simply detect whether the output file exists.\\
\begin{lstlisting}[language=Python]
def test_task_141(trajectory):
    output_folder = "./output/141"
    html_files_exist = any(file.endswith('.html') for file in os.listdir(output_folder))
    assert html_files_exist
\end{lstlisting}

\section{Code as Action VS Function Calling}
\label{CodeVsFunction}
We conducted an ablation study on the format of calling code interpreter. For function calling, we defined an \textit{execute\_python} function where the LLM passes a code string as the parameter and uses a special token \textit{<|tool\_call|>} before the function.

We repurposed all code snippets in PyInstruct, CodeFeedback, and CodeActInstruct to follow the function calling format. We trained the model without continue pre-train on code-rich corpus for it exactly aligns with our format. Compare it to PyLlama3(w/o cpt), Table \ref{Compare format} shows that the function calling format struggles with real-world code tasks. The model often fails to follow the format and frequently neglects to call the \textit{execute\_python} tool, leading to failures.

\begin{table}[h]
\centering
\begin{tabular}{p{3.25cm}p{1cm}p{1cm}p{1cm}}
\hline
\textbf{Format} &  \textbf{Pass Rate(UT)} & \textbf{Average Turns} \\
\hline
Function call & 19.6 & 8.3  \\
Ours & 58.7 & 6.3   \\
\hline
\end{tabular}
\caption{Function Call vs Our format}
\label{Compare format}
\end{table}

\section{Compare LLM Evaluator and Unit Test}
\label{LLMVSUT}
\begin{table}[h]
\centering
\begin{tabular}{p{1cm}p{1cm}p{1cm}}
\hline
\textbf{Good} &  \textbf{Same} & \textbf{Bad} \\
\hline
12 & 43 & 5 \\
\hline
\end{tabular}
\caption{Compare the LLM Evaluator with the Unit Test. "Good" indicates that the LLM Evaluator is correct while the Unit Test is incorrect. "Same" means both the LLM Evaluator and the Unit Test provide the same pass-or-fail decision. "Bad" signifies that the LLM Evaluator is incorrect while the Unit Test is correct.}
\label{GSB}
\end{table}

\begin{figure*}[h]
    \centering
    \includegraphics[width=1\linewidth]{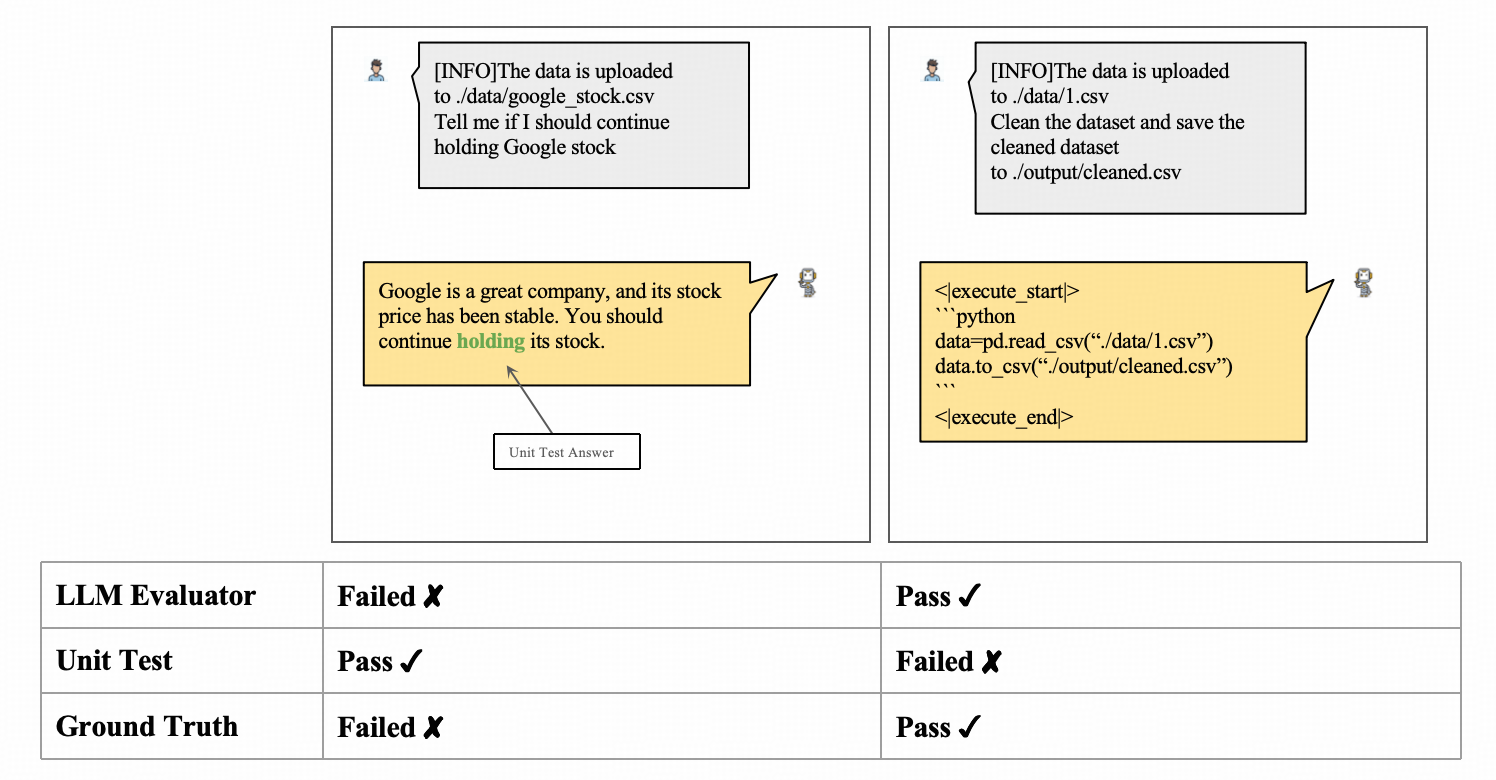}
    \caption{Compare Unit Test and LLM Evaluator}
    \label{UTVSLLM}
\end{figure*}

We randomly select 60 samples from all the generated trajectories and judge them manually. Table \ref{GSB} shows in most situations, LLM Evaluator and Unit Test reach a consensus. In other situations, LLM Evaluator and Unit Test have their own advantages and drawbacks. Figure \ref{UTVSLLM} shows two examples.




\section{Task Distribution}
Table \ref{Distribution} shows the number of each type of task in PyBench.
\begin{table*}[h]
\centering
\begin{tabular}{lllll}
\hline
\textbf{Chart.} & \textbf{Text.} & \textbf{Image.} & \textbf{Math.} & \textbf{Software.} \\
\hline
62 & 23 & 48 & 6 & 4 \\
\hline
\end{tabular}
\caption{The distribution of each type of task in PyBench.}
\label{Distribution}
\end{table*}

\section{Python Packages in our environment}
\label{PythonPackages}
\begin{quote}
\begin{Verbatim}[fontsize=\small]
absl-py
analytics-python
attrs
audioread
beautifulsoup4
bokeh
cachetools
CairoSVG
cairosvg
chardet
click
click-plugins
compressed-rtf
debugpy
decorator
defusedxml
deprecat
docx2txt
ebooklib
email-validator
fastapi
fastjsonschema
fbprophet
ffmpeg-python
ffmpy
fire
fitz
Flask
flask
Flask-CacheBuster
flask-cachebuster
Flask-Cors
flask-cors
Flask-Login
flask-login
fonttools
frontend
fpdf
future
fuzzywuzzy
gensim==3.8.2
gradio
graphviz
h5py
html5lib
httpx
IMAPClient
imageio
imageio-ffmpeg
imgkit
ipython
Jinja2
jinja2
jieba
json5
jsonpickle
jsonschema
jupyter-client
jupyter-core
jupyter-server
jupyterlab
jupyterlab-pygments
jupyterlab-server
keras
langchain
langchain-experimental
librosa
loguru
lxml
markdown2
markdownify
MarkupSafe
matplotlib
matplotlib-inline
matplotlib-venn
moviepy
murmurhash
nbclient
nbconvert
nbformat
networkx
nltk
notebook
numba
numexpr
numpy
numpy-financial
openai
opencv-python
openpyxl
orjson
pandas
pdf2image
pdfkit
pdfminer.six
pdfplumber
Pillow
pillow
plotly
psutil
PyAudio
PyMuPDF
pydantic
pydub
Pygments
pygments
pylog
PyMuPDF
pymupdf
pypandoc
pyparsing
PyPDF2
pypdf2==2.12.0
pytesseract
pytest
python-dateutil
python-docx
python-multipart
python-pptx
pytz
PyWavelets
pywavelets
pygame
PyYAML
pyyaml
qrcode
rarfile
requests
scikit-image
scikit-learn
scipy
seaborn
sentencepiece
Shapely
soundfile
SoundFile
SpeechRecognition
speechrecognition
starlette
statsmodels
svglib
svgwrite
sympy
sumy
tabula
tabulate
textblob
tifffile
toml
torch
torchaudio
torchtext
torchvision
tornado
tqdm
typing-extensions
tzdata
tzlocal
ujson
urllib3
uvicorn
Wand
wand
websocket-client
websockets
Werkzeug
werkzeug
wordcloud
xgboost
xlrd
XlsxWriter
xlsxwriter
xml-python
zipp

\end{Verbatim}
\end{quote}